\documentclass{article}

\usepackage{epsfig}
\usepackage{amsmath}
\usepackage{amsthm}
\usepackage{amssymb}
\usepackage{eepic}
\usepackage[T1]{fontenc}

\newcommand{\R}{\mathbb{R}} %
\newcommand{\T}{\mathbb{T}} %
\newcommand{\Z}{\mathbb{Z}} %
\newcommand{\N}{\mathbb{N}} %
\newcommand{\hmax}{{h_{\max}}}
\newcommand{\eqLaw}{{\stackrel{\mathcal L}{=}}}
\newcommand{\ep}{{\varepsilon}} %

\newcommand{\suchthat}{, } %
\newcommand{\set}[2]{\left\{#1\suchthat#2\right\}} %
\newcommand{\abs}[1]{{\left\lvert#1\right\rvert}} %
\newcommand{\paren}[1]{{\left( #1 \right)}} %
\newcommand{\deq}{:=} 
\newcommand{\card}{\#} 
\newcommand{\bsb}[1]{\boldsymbol{#1}} 

\newcommand{\brho}{\bsb{\rho}}
\newcommand{\bhmin}{{h_{\min}}}
\newcommand{\bhmax}{{h_{\max}}}
\newcommand{\Lebesgue}{\mathcal{L}}

\newtheorem{defi}{Definition} %
\newtheorem{theo}{Theorem} %
\newtheorem{prop}{Proposition}[section] %

\begin{document}

\title{Random Wavelet Series:\\ Theory and  Applications}

\newcommand{\labname}{Laboratoire d'Analyse et de Mathématiques Appliquées}
\newcommand{\univname}{Université Paris-XII Val de Marne}

\author{Jean-Marie Aubry and Stéphane Jaffard\thanks{Also supported by the Institut Universitaire de France}\\ Laboratoire d'Analyse et de Mathématiques Appliquées\\ Université Paris \textsc{xii}-Val de Marne\\ UMR CNRS 8050}

\maketitle

\begin{abstract}
  Random Wavelet Series form a  class of random
  processes with multifractal properties.  We give three
  applications of this construction. %
  First, we synthesize a random function having any given spectrum of
  singularities satisfying some conditions
  (but including non-concave spectra). %
  Second, these processes provide  examples where the multifractal
  spectrum coincides with the spectrum of large deviations, and we
  show how to recover it  numerically. %
  Finally,  particular cases of these processes satisfy a generalized
  selfsimilarity relation proposed in the theory of fully developed
  turbulence.
\end{abstract}

\section{Introduction: Random Wavelet Series}
\label{sec:intr-rand-wavel}

The story of random Fourier series largely coincides with the
development of harmonic analysis in the twentieth century, starting
with the pioneering work of Borel, later developed by Wiener, Salem
and Zygmund, to quote but a few, and culminating with the famous book
of Kahane~\cite{kahane85:_some} which made this theory accessible to a
large audience.  Since the mid 80's, wavelet bases have proved to be a
preponderant alternative option to the trigonometric system, in order
to analyze and synthesize functions and signals. Therefore, a very
natural problem is to study random wavelet series.  Surprisingly, it
turns out that such series have properties that differ widely from
those of random Fourier series.  This difference is particularly
striking when one considers pointwise Hölder regularity.  Recall
that, if $\alpha >0$, a function $f $, defined on $\R$ is $C^\alpha
(x_0 )$ if there exists a polynomial $P$ of degree at most $[ \alpha
]$ such that
\begin{equation}
  \label{eq:9}
  | f(x) -P(x-x_0 )| \leq C | x-x_0 |^\alpha .
\end{equation}
The Hölder exponent of $f$ at $x_0$, denoted by $h_f (x_0)$, is
defined as
\begin{equation*}
  h_f (x_0) \deq \sup \{ \alpha: f \in C^\alpha (x_0 ) \}.
\end{equation*}
Under very general assumptions, random Fourier series with independent
coefficient have {\em everywhere} the same Hölder exponent.  (The
differences that appear in the modulus of continuity at different
points and allow to draw a difference between slow points and fast
points are logarithmic corrections in the right hand side
of~(\ref{eq:9}), of the form $\log (| x-x_0| )^\gamma$; compare for
instance in \cite{kahane85:_some} the results concerning the uniform
modulus of continuity given by Theorem 2 in Chapter 7 with the results
concerning irregularity everywhere in Section 6 of Chapter 8.)

By contrast, random wavelet series with independent coefficients have
a Hölder exponent which is a highly irregular random function: Let
\begin{equation*}
  E_h  \deq \set{x}{f \text{ has Hölder exponent
    } h \text{ at } x};
\end{equation*}
if the distributions of the wavelet coefficients depend only on the
scale of the wavelet (and some mild additional hypotheses), the sets
$E_h$ are non-empty when $h $ takes values in an interval of non-empty
interior $[h_{min}, h_{max}]$, in which case the $E_h$ are random
fractal sets, see \cite{aubry:_gener_random_wavel_series}. The
Hausdorff dimension of $E_h$, which is denoted by $d(h)$ is called the
{ \em spectrum of singularities} of the sample path (we use the
traditional convention $\dim_H(\emptyset) \deq -\infty$).
In this paper we will recall the previous results concerning random
wavelet series, and continue this study by showing in particular that
the model supplied by random wavelet series is compatible with several
turbulence models that have been proposed in the past.

There are many real-life situations where, although the independence
condition is not necessarily satisfied, good estimates or models are
known on the distribution of wavelet coefficients. Let us mention a
few examples:
\begin{itemize}
\item Several authors (Buccigrossi \emph{et al.} \cite{BS}, Huang
  \emph{et al.} \cite{HM}, Mallat \cite{Ma}, Simoncelli \cite{Sim},
  Vidakovic \cite{Vi1}) have studied the statistics of the wavelet
  coefficients of large collections of natural images and observed
  that these statistics are highly non-Gaussian.  Exponential power
  distributions (of density $C e^{ -A |x|^{\alpha}}$) fit very well
  these statistics.
  
\item Cascade-type models for the evolution of the probability density
  function of the wavelet coefficients through the scales have been
  proposed to model the velocity in the context of fully developed
  turbulence (these models were initially proposed by Castaing
  \emph{et al.}  \cite{chilla96:_multip} for the increments of the
  velocity, and then fitted to the wavelet setting by Arneodo \emph{et
    al.}  \cite{arneodo98:_random}).  Random multiplicative models
  have also been considered in statistics, see Vidakovic \cite{Vi2}.
  
\item Bayesian inference techniques based on a priori models for the
  distributions of wavelet coefficients at each scale have been widely
  studied (see for instance Abramovich \emph{et al.}
  \cite{abramovich97:_wavel_bayes}, Johnstone \cite{JS}, Müller
  \emph{et al.}  \cite{MV}) to improve the usual wavelet-based
  denoising algorithms by using the additional information supplied by
  the distributions of wavelet coefficients (instead of using only the
  Besov regularity for instance).
  
\item In multifractal analysis, several formulas (Lévy-Vehel \emph{et
    al.}  \cite{LR}, Evertsz \emph{et al.}  \cite{EM}, Meneveau
  \emph{et al.}  \cite{MS}, Riedi \cite{Rie}) were also proposed in
  order to derive spectra of singularities from distributions of
  increments of the function. These formulas are referred to as { \sl
    large deviation multifractal formalisms}; they can be easily
  extended to a wavelet setting.
\end{itemize}

We will  explore how random wavelet series can play a role in
some of these models.  Let us first recall their definition and main
properties. We consider functions on $\T \deq \R/\Z$ (1-periodic
functions).  Let $\psi$ be a mother wavelet such that the periodized
wavelet family
\begin{equation*}
  \set{\psi_{j, k}: x \mapsto \sum_{l \in \Z} \psi(2^j (x-l) -
    k)}{j\in\N, 0\leq k < 2^j}
\end{equation*}
form, together with the constant function $x \mapsto 1$, an orthogonal
basis of $L^2(\mathbb T)$ (but not orthonormal, the $L^\infty$
normalization being more convenient for our purpose). The wavelet
coefficients of a function $f$ are
\[  C_{j, k} =  2^j \int_0^1  f(x) \psi_{j, k} (x)  d x. \]
In the following, we will assume that the wavelet $\psi$ belongs to
the Schwartz class, which will simplify the statements of the main
results.  Note however that the regularity results that we will state
remain partly valid when using a wavelet with limited regularity: In
this case, one has to assume that the maximal Hölder regularity of
the process (that will be denoted by $h_{max}$) is strictly smaller
than the uniform Hölder regularity of the wavelet.

\subsection{Random wavelet coefficients}
\label{sec:rand-wavel-coeff}

\begin{defi}
  \label{defi:rws}
  A periodic distribution $f$ is a random wavelet series (RWS) if
  its wavelet coefficients $C_{j, k}$ in the basis above satisfy the
  following requirements:
  \begin{enumerate}
  \item $\forall j$, the $C_{j, k}$ ($ k \in \{ 0, \dots , 2^j-1$\}) are
    identically distributed random variables; the probability
    distribution of $-\frac{\log_2(\abs{C_{j, k}})}{j}$ is denoted by
    $\brho_j$; it is defined on $\R \cup \{+ \infty\}$;
  \item the $C_{j, k}$  ($j \in \N$, $  k \in \{0, \dots , 2^j-1\}$)  are
    independent;
  \item there exists $\gamma > 0$ such that
    \begin{equation}
      \label{eq:1}
      \brho(\alpha) \deq \inf_{\ep > 0} \limsup_{j \rightarrow
        +\infty} \frac{ \log_2 \left( 2^j
          \brho_j([\alpha  -\ep, \alpha +\ep])
        \right) }{j}
    \end{equation}
    is strictly negative for $\alpha < \gamma$.
  \end{enumerate}
\end{defi}

The third requirement is necessary in order to be sure that the series
$\sum c_{j,k} \psi_{j,k}$ is convergent in the sense of distributions.
The function $\brho$ thus defined is called the \emph{upper
  logarithmic density} of the process. It is upper semi-continuous,
but not necessarily monotonous. We do not make any other assumption on
the probability measures $\brho_j$; note that $\brho_j(\{+\infty\})$
is simply the probability that $C_{j, k}=0$.

\subsection{Histograms of wavelet coefficients}
\label{sec:wavel-coeff-hist}

Consider now an arbitrary function (or distribution) $f$, which, for
instance,  can be  a sample path  of a stochastic process.  Let
$N_j(\alpha)
\deq
\card \set{k}{ |
  C_{j, k}| \geq 2^{ - \alpha j}}$, and
\begin{equation}
  \label{eq:2}
  \rho(\alpha) \deq \inf_{\ep > 0} \limsup_{j \to + \infty}
  \frac{\log_2(N_j(\alpha + \ep) - N_j(\alpha - \ep))}{j}.
\end{equation}

Without any additional assumption, we expect $\rho$ to be a random
function, whereas, if $f$ is a RWS, $\brho$ is deterministic. A
first result of~\cite{aubry:_gener_random_wavel_series} links these
two functions.
\begin{theo}
  \label{theo:brhorho}
  Let $f$ be a  RWS, and
  \begin{equation*}
    W \deq \set{\alpha}{\forall \epsilon > 0, \sum_{j\in\N} 2^j
      \brho_j([\alpha-\epsilon, \alpha+\epsilon]) = +\infty}.
  \end{equation*}
  With probability one, for all $\alpha$,
  \begin{equation*}
    \rho(\alpha) =
    \begin{cases} \brho(\alpha) \text{ if } \alpha \in W
      ; \\ -\infty \text{ else.}
    \end{cases}
  \end{equation*}
\end{theo}
One immediately checks that $\brho(\alpha) > 0 \Rightarrow \alpha\in
W$ and $\brho(\alpha) < 0 \Rightarrow \alpha \not \in W$. If
$\brho(\alpha)=0$, then $\rho(\alpha)$ can be either (almost surely) $0$ or
$-\infty$.

\subsection{Spectrum of singularities}
\label{sec:spectr-sing}

A second result gives the spectrum of singularities for a RWS
Naturally if $W=\emptyset$, then $f$ is almost surely globally
$C^\infty$.  Otherwise, let us define
\begin{equation*}
  \bhmin \deq \inf(W),
\end{equation*}
and, assuming that $\exists \alpha$, $\brho(\alpha) > 0$,
\begin{equation*}
  \bhmax \deq \paren{\sup_{\alpha > 0}
    \frac{\brho(\alpha)}{\alpha}}^{-1}.
\end{equation*}

Note that, because $\brho$ is upper semi-continuous, $\brho(\bhmin)
\geq 0$.

\begin{figure}[htbp]
  \begin{center}
    \setlength{\unitlength}{0.1pt}
\begin{picture}(3000,1800)(0,0)
\footnotesize
\thicklines \path(370,249)(411,249)
\thicklines \path(2876,249)(2835,249)
\put(329,249){\makebox(0,0)[r]{ 0}}
\thicklines \path(370,494)(411,494)
\thicklines \path(2876,494)(2835,494)
\put(329,494){\makebox(0,0)[r]{ 0.2}}
\thicklines \path(370,739)(411,739)
\thicklines \path(2876,739)(2835,739)
\put(329,739){\makebox(0,0)[r]{ 0.4}}
\thicklines \path(370,984)(411,984)
\thicklines \path(2876,984)(2835,984)
\put(329,984){\makebox(0,0)[r]{ 0.6}}
\thicklines \path(370,1228)(411,1228)
\thicklines \path(2876,1228)(2835,1228)
\put(329,1228){\makebox(0,0)[r]{ 0.8}}
\thicklines \path(370,1473)(411,1473)
\thicklines \path(2876,1473)(2835,1473)
\put(329,1473){\makebox(0,0)[r]{ 1}}
\thicklines \path(370,1718)(411,1718)
\thicklines \path(2876,1718)(2835,1718)
\put(329,1718){\makebox(0,0)[r]{ 1.2}}
\thicklines \path(370,249)(370,290)
\thicklines \path(370,1718)(370,1677)
\put(370,166){\makebox(0,0){ 0}}
\thicklines \path(871,249)(871,290)
\thicklines \path(871,1718)(871,1677)
\put(871,166){\makebox(0,0){ 0.5}}
\thicklines \path(1372,249)(1372,290)
\thicklines \path(1372,1718)(1372,1677)
\put(1372,166){\makebox(0,0){ 1}}
\thicklines \path(1874,249)(1874,290)
\thicklines \path(1874,1718)(1874,1677)
\put(1874,166){\makebox(0,0){ 1.5}}
\thicklines \path(2375,249)(2375,290)
\thicklines \path(2375,1718)(2375,1677)
\put(2375,166){\makebox(0,0){ 2}}
\thicklines \path(2876,249)(2876,290)
\thicklines \path(2876,1718)(2876,1677)
\put(2876,166){\makebox(0,0){ 2.5}}
\thicklines \path(370,249)(2876,249)(2876,1718)(370,1718)(370,249)
\put(-200,983){\makebox(0,0)[l]{$d(h)$}}
\put(1623,42){\makebox(0,0){$h$}}

\thicklines

\path(461,249)(471,294)(497,392)(522,480)(547,560)(573,634)(598,703)(623,767)(648,826)(674,881)(699,934)(955,1467)

\thinlines

\dottedline{60}(370,249)(699,934)

\path(724,981)(750,1025)(775,1067)(800,1106)(826,1143)(851,1177)(876,1208)(902,1238)(927,1265)(952,1290)(978,1313)(1003,1335)(1028,1354)(1053,1372)(1079,1389)(1104,1403)(1129,1416)(1155,1428)(1180,1438)(1205,1447)(1231,1455)(1256,1461)(1281,1466)(1307,1469)(1332,1472)(1357,1473)(1383,1473)(1408,1472)(1433,1470)(1458,1467)(1484,1463)(1509,1457)(1534,1451)(1560,1444)(1585,1436)(1610,1427)(1636,1417)(1661,1406)(1686,1394)(1712,1382)

\path(1712,1382)(1737,1368)(1762,1354)(1788,1339)(1813,1323)(1838,1307)(1863,1289)(1889,1271)(1914,1252)(1939,1233)(1965,1212)(1990,1191)(2015,1169)(2041,1147)(2066,1124)(2091,1100)(2117,1076)(2142,1051)(2167,1025)(2193,998)(2218,971)(2243,944)(2268,916)(2294,887)(2319,858)(2344,828)(2370,797)(2395,766)(2420,734)(2446,702)(2471,669)(2496,636)(2522,602)(2547,568)(2572,533)(2598,498)(2623,462)(2648,426)(2673,389)(2699,351)(2724,314)(2749,275)(2767,249)

\dottedline{60}(370,1473)(2876,1473)

\end{picture}
    \caption{Upper logarithmic density (thin curve) and
      spectrum of singularities (bold curve) for the Poisson kernel ($c=1$,
      $\alpha_0=0$).}
    \label{fig:poisson}
  \end{center}
\end{figure}
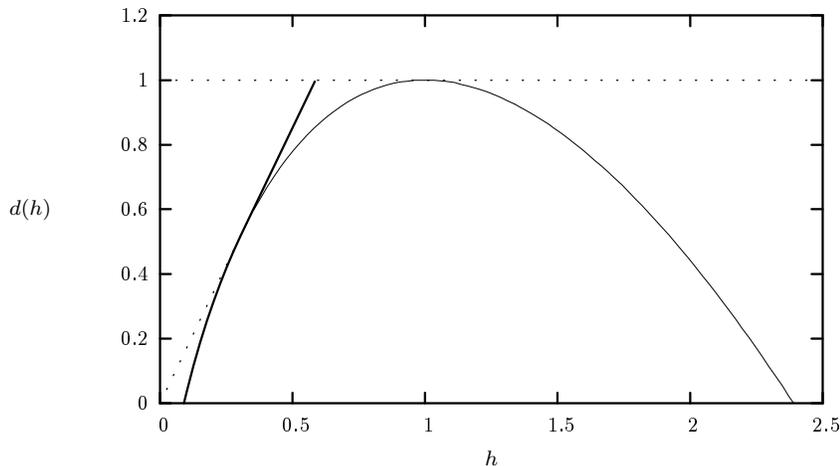

\begin{theo}
  \label{theo:rwsspectrum}
  Let $f$ be a RWS  Almost-surely,
  \begin{itemize}
  \item the almost-everywhere Hölder exponent is $\hmax$;
  \item for all $h \in [h_{\min}, \hmax]$, $d(h) = h \sup_{\alpha
      \in (0,h]} \frac{\bsb\rho(\alpha)}{\alpha}$;
  \item $d(h) = -\infty$ else.
  \end{itemize}
\end{theo}
See Figure~\ref{fig:poisson} for an illustration.

In \cite{aubry:_gener_random_wavel_series} it was implicitly assumed
that there exists $\alpha$ such that $\brho(\alpha) > 0$.  However, it
may happen that this condition does not hold, and yet $W$ is not
empty. Consider for instance the RWS defined by fixing $\alpha_0 > 0$,
and $\forall j \in \N$, $ \brho_j(\{\alpha_0\}) = j 2^{-j}$ and $
\brho_j(\{+\infty\}) = 1 - j 2^{-j}$ (here $W = \{\alpha_0\}$). In
that (degenerate) case, we get an almost sure ``flat'' spectrum.
\begin{prop}
  Let $f$ be a RWS such that $\forall \alpha$, $\brho(\alpha) \leq
  0$, but $W \neq \emptyset$. Define $\bhmin$ as above. Then, almost
  surely,
  \begin{itemize}
  \item the almost-everywhere Hölder exponent is $+\infty$;
  \item for all $h \geq \bhmin$, $d(h) = 0$;
  \item $d(h) = -\infty$ else.
  \end{itemize}
\end{prop}

\begin{proof}
  First, according to Theorem \ref{theo:brhorho}, $\brho(\gamma) < 0$
  implies a. s. $\rho(\gamma) = -\infty$, which means that
  with at most finitely many exceptions, $\abs{C_{j,k}} < 2^{-\gamma
    j}$.

  For $\alpha \geq 0$ let $K^j(\alpha)\deq \set{k}{\abs{C_{j,k}} \geq
    2^{-\alpha j}}$ and if $d \leq 1$, let
  \begin{equation*}
    E^j(\alpha,d)\deq \bigcup_{k\in K^j(\alpha)} (k2^{-j} - 2^{-d j},
    k2^{-j} + 2^{-d j}),
  \end{equation*}
  and
  \begin{equation*}
    E(\alpha,d) \deq \limsup_{j\to+\infty} E^j(\alpha,d).
  \end{equation*}
  Note that $E(\alpha,d)$ is increasing in $\alpha$ and decreasing in $d$.
  Because $\brho(\alpha) \leq 0$, $\forall d > 0$, by the Borel-Cantelli
  lemma, almost surely
  $E(\alpha,d)$ has Lebesgue measure $\Lebesgue(E(\alpha,d)) =
  0$. Then, with
  \begin{equation*}
    E \deq \bigcup_{m \geq 1} E\paren{m,\frac\gamma m},
  \end{equation*}
  almost surely $\Lebesgue(E) = 0$. If $x \not \in E$, then for all $m
  \geq 1$, for all $j, k$ with at most a finite number of exceptions,
  either $\abs{C_{j,k}} < 2^{-m j} $ or $\abs{x-k 2^{-j}} \geq
  2^{-\frac{j \gamma}{m} }$, in which case $\abs{C_{j,k}} <
  \abs{x-k 2^{-j}}^m$. Using the classical wavelet characterization of
  pointwise regularity, this proves that $f \in C^\infty(x)$.

  The last two points are similar to
  Theorem~\ref{theo:rwsspectrum} (in the case where  $\brho(\alpha) >
  0$ happens only for some $\alpha > h_0 > \bhmin$).

\end{proof}

To conclude this overview, let us mention the following result, also proved in
\cite{aubry:_gener_random_wavel_series}.  A function $f$ is called a
uniform Hölder function if there exists $\epsilon > 0$ such that $f
\in C^\epsilon(\T)$.

\begin{prop}
  The spectrum of
  singularities of any uniform Hölder function $f$ satisfies the
  inequality
  \begin{equation}
    \label{eq:3}
    d(h) \leq h \sup_{\alpha \in (0,h]} \frac{\rho(\alpha)}{\alpha}.
  \end{equation}
\end{prop}

\section{Synthesis of multifractal processes}
\label{sec:synth-mult-proc}

When a multifractal model is proposed, a natural step is to construct
a function or a random process having a given spectrum of
singularities $d(h)$.  According to Theorem~\ref{theo:rwsspectrum},
the candidates for RWS spectra are right-continuous functions
satisfying the following properties: $d \leq 1$; $d \geq 0$ on an
interval $[ h_{min}, h_{max} ]$ where the function $h \mapsto
\frac{d(h)}{h}$ is increasing; outside this interval $d(h) = -\infty$;
and $d (h_{max}) = 1$.  These conditions are
also sufficient.

\begin{prop}
  Let $d(h)$ satisfy the conditions listed above. Take, for all $j \in
  \N$, for all $\alpha \in \R$,
  \begin{equation*}
    \brho_j(d \alpha) \deq \frac{j \ln(2)}{\hmax} 2^{j(d(\alpha)-1)} d\alpha
  \end{equation*}
  and $\brho_j(\{+\infty\}) \deq 1 - \int_0^{\hmax} \brho_j(d
  \alpha)$.  Then $\brho_j$ is a probability measure on $\R \cup \{ +
  \infty \}$, and its upper logarithmic density is $\brho(\alpha) =
  d(\alpha)$.
\end{prop}

\begin{proof}
  Note that $d(\alpha) \leq \frac\alpha\hmax$. It follows that
  \begin{align*}
    \int_\R \brho_j(d \alpha) &= \int_0^{\hmax} \brho_j(d \alpha) \\
    &\leq \frac{j \ln(2)}{\hmax}
    2^{-j} \int_0^{\hmax} 2^{j \frac\alpha\hmax} d\alpha \\
    &\leq 1,
  \end{align*}
  which ensures, together with the definition of
  $\brho_j(\{+\infty\})$, that $\brho_j$ is a probability measure on
  $\R \cup \{ + \infty \}$.

  Let us now compute the upper logarithmic density. We have
  \begin{gather*}
    \frac{ \log_2 \paren{\ep \frac{j \ln(2)}{\hmax} }}{j}+ d(\alpha)
    \leq \frac{ \log_2 \paren{ 2^j \int_{\alpha -\ep}^{\alpha +\ep}
        \brho_j(t) d t } }{j} \leq \frac{ \log_2
      \paren{2 \ep \frac{j \ln(2)}{\hmax} } }{j} + d(\alpha+\ep), \\
    d(\alpha) \leq \limsup_{j \to + \infty} \frac{ \log_2 \paren{ 2^j
        \int_{\alpha -\ep}^{\alpha +\ep} \brho_j(t) d t } }{j} \leq
    d(\alpha+\ep)
  \end{gather*}
  hence, letting $\ep \to 0$ and using right-continuity,
  $\brho(\alpha) = d(\alpha)$.
\end{proof}

The almost sure spectrum of singularities of the corresponding
RWSis then $d$.  To synthesize it, for all $j \in \N$ we draw
independently $2^j$ random variables $\alpha_{j k}$ with law $\brho_j$
(using the rejection method if necessary), and let $C_{j, k} \deq
\chi_{j k} 2^{-j \alpha_{j k}}$ ($\chi_{j k}$ is an arbitrary sign or
phase). Then
\begin{equation}
  \label{eq:4}
  f(x) = \sum_{j\in\N} \sum_{k=0}^{2^j-1} C_{j, k} \psi_{j, k}(x)
\end{equation}
is a realization of the process.

\begin{figure}[htbp]
  \begin{center}
    \psfig{angle=270,width=0.8\textwidth,file=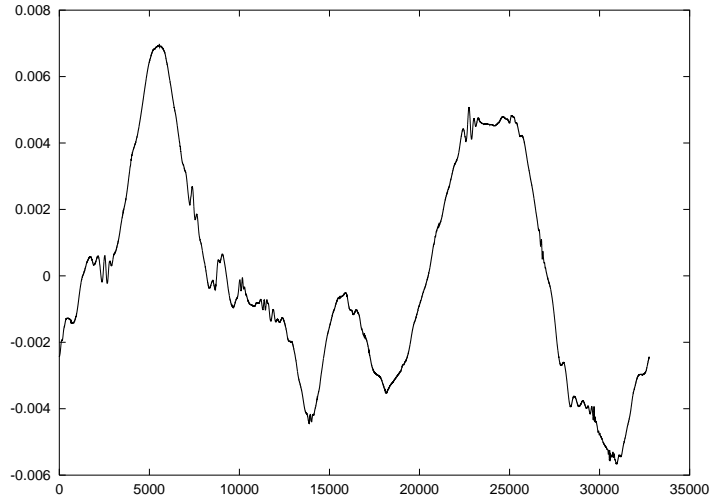}
    \caption{A multifractal function with non-concave spectrum}
    \label{fig:mulpro}
  \end{center}
\end{figure}

This allows in particular to show processes with non-concave spectra.
An example is given on Figure~\ref{fig:mulpro} with $d(h) \deq (h -
\frac12)^2$ on $[\frac12, \frac32]$ and $-\infty$ elsewhere.  The
wavelet used is Daubechies 10 (extremal phase).

\section{Spectrum of large deviation}
\label{sec:mult-form}

For a function given in a sampled form (signals, images), one often
wishes to compute its spectrum of singularity. Just applying the
definition (identifying the sets of common Hölder exponent and
computing their Hausdorff dimension) is clearly not feasible, so
numerical procedures called \emph{multifractal formalisms} have been
proposed instead. In general, a multifractal formalism does not yield
the correct spectrum, but an upper bound, and is proved valid only on
certain functions. For instance, the ``classical'' multifractal
formalism, derived from the original ideas of Parisi and Frisch
\cite{frisch}, consists in computing the so-called structure function
\begin{equation*}
  \tau(q) \deq \liminf_{j\to+\infty} \frac{\log_2\paren{\sum_k
      \abs{C_{j, k}}^q}}{-j},
\end{equation*}
and taking its Legendre transform
\begin{equation}
  \label{eq:5}
  d_1(h) \deq \inf_{q } h q - \tau(q)
\end{equation}
as an estimation for $d(h)$.  It was proved by Jaffard~\cite{jaffard1}
that for any function $f$, if $q_c$ is the (only) solution to
$\tau(q_{c}) = 0$, and if the infimum in (\ref{eq:5}) is taken for ${q
  \geq q_c}$,  then
$d(h) \leq d_1(h)$; moreover, equality in~(\ref{eq:5}) holds for
selfsimilar functions (Jaffard
\cite{jaffard2}). Note, however, that this cannot be true for all
functions; in particular the right-hand term of~(\ref{eq:5}) is
concave, whereas in general the spectrum is not.

One can also prove that, for any function (even for a tempered
distribution), for all $h$,
\begin{equation}
  \label{eq:6}
  h \sup_{\alpha \in (0,h]} \frac{\rho(\alpha)}{\alpha} \leq \inf_{q
    \geq q_c} h q - \tau(q);
\end{equation}
this implies that~(\ref{eq:3}) is sharper than~(\ref{eq:5}).
Actually,
\begin{equation*}
  d_2(h) \deq h \sup_{\alpha \in (0,h]} \frac{\rho(\alpha)}{\alpha}
\end{equation*}
is equivalent to the \emph{spectrum of large deviations} of~\cite{LR};
we know from Theorem~\ref{theo:rwsspectrum} that for RWS it
coincides with the spectrum of singularities.

Because of~(\ref{eq:6}), this new multifractal formalism will be valid
whenever the classical one is valid; moreover, since the spectrum thus
obtained is not necessary concave, its domain of validity is strictly
larger than the classical one. %
Note that $\rho(\alpha)$ may not be easy to compute numerically,
because~(\ref{eq:2}) involves a double limit. But if we define
\begin{equation*}
  \lambda(\alpha) \deq \limsup_{j\to+\infty}
  \frac{\log_2(N_j(\alpha))}{j},
\end{equation*}
which is increasing, we can show that its upper closure $\bar\lambda$
(whose hypograph is the closure of the hypograph of $\lambda$)
satisfies $\bar\lambda(\alpha) = \sup_{\alpha' \leq \alpha}
\rho(\alpha')$. Hence $d_2(h) = \sup_{\alpha \in (0,h]}
\frac{\bar\lambda(\alpha)}{\alpha}$ as well, which is easier to
compute.

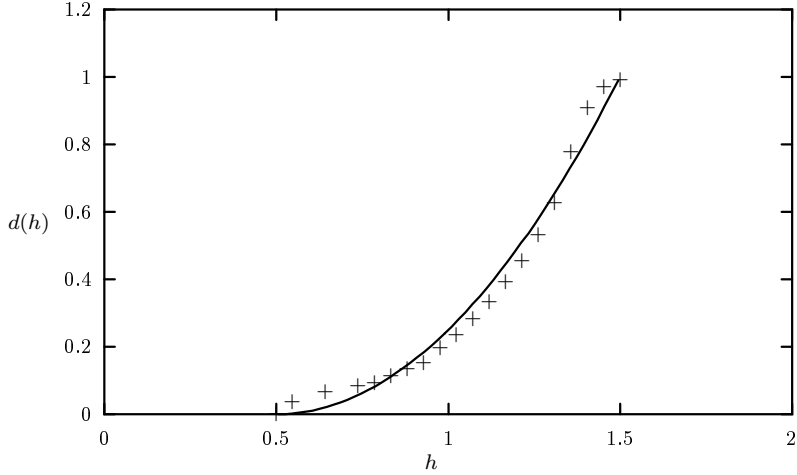
\begin{figure}[htbp]
  \begin{center}
    \setlength{\unitlength}{0.2pt}
\newcommand{\markassin}{+}
\begin{picture}(1500,900)(0,0)
\footnotesize
\thicklines \path(132,90)(152,90)
\thicklines \path(1433,90)(1413,90)
\put(110,90){\makebox(0,0)[r]{0}}
\thicklines \path(132,218)(152,218)
\thicklines \path(1433,218)(1413,218)
\put(110,218){\makebox(0,0)[r]{0.2}}
\thicklines \path(132,345)(152,345)
\thicklines \path(1433,345)(1413,345)
\put(110,345){\makebox(0,0)[r]{0.4}}
\thicklines \path(132,473)(152,473)
\thicklines \path(1433,473)(1413,473)
\put(110,473){\makebox(0,0)[r]{0.6}}
\thicklines \path(132,601)(152,601)
\thicklines \path(1433,601)(1413,601)
\put(110,601){\makebox(0,0)[r]{0.8}}
\thicklines \path(132,728)(152,728)
\thicklines \path(1433,728)(1413,728)
\put(110,728){\makebox(0,0)[r]{1}}
\thicklines \path(132,856)(152,856)
\thicklines \path(1433,856)(1413,856)
\put(110,856){\makebox(0,0)[r]{1.2}}
\thicklines \path(132,90)(132,110)
\thicklines \path(132,856)(132,836)
\put(132,45){\makebox(0,0){0}}
\thicklines \path(457,90)(457,110)
\thicklines \path(457,856)(457,836)
\put(457,45){\makebox(0,0){0.5}}
\thicklines \path(783,90)(783,110)
\thicklines \path(783,856)(783,836)
\put(783,45){\makebox(0,0){1}}
\thicklines \path(1108,90)(1108,110)
\thicklines \path(1108,856)(1108,836)
\put(1108,45){\makebox(0,0){1.5}}
\thicklines \path(1433,90)(1433,110)
\thicklines \path(1433,856)(1433,836)
\put(1433,45){\makebox(0,0){2}}
\put(-50,450){\makebox(0,0)[l]{\shortstack{$d(h)$}}}
\put(750,0){\makebox(0,0){$h$}}
\thicklines \path(132,90)(1433,90)(1433,856)(132,856)(132,90)
\put(457,90){\raisebox{0pt}{\makebox(0,0){$\markassin$}}}
\put(488,114){\raisebox{0pt}{\makebox(0,0){$\markassin$}}}
\put(550,133){\raisebox{0pt}{\makebox(0,0){$\markassin$}}}
\put(612,144){\raisebox{0pt}{\makebox(0,0){$\markassin$}}}
\put(643,149){\raisebox{0pt}{\makebox(0,0){$\markassin$}}}
\put(674,163){\raisebox{0pt}{\makebox(0,0){$\markassin$}}}
\put(705,175){\raisebox{0pt}{\makebox(0,0){$\markassin$}}}
\put(736,188){\raisebox{0pt}{\makebox(0,0){$\markassin$}}}
\put(767,216){\raisebox{0pt}{\makebox(0,0){$\markassin$}}}
\put(798,241){\raisebox{0pt}{\makebox(0,0){$\markassin$}}}
\put(829,270){\raisebox{0pt}{\makebox(0,0){$\markassin$}}}
\put(860,303){\raisebox{0pt}{\makebox(0,0){$\markassin$}}}
\put(891,340){\raisebox{0pt}{\makebox(0,0){$\markassin$}}}
\put(922,380){\raisebox{0pt}{\makebox(0,0){$\markassin$}}}
\put(953,429){\raisebox{0pt}{\makebox(0,0){$\markassin$}}}
\put(984,491){\raisebox{0pt}{\makebox(0,0){$\markassin$}}}
\put(1015,588){\raisebox{0pt}{\makebox(0,0){$\markassin$}}}
\put(1046,671){\raisebox{0pt}{\makebox(0,0){$\markassin$}}}
\put(1077,710){\raisebox{0pt}{\makebox(0,0){$\markassin$}}}
\put(1108,724){\raisebox{0pt}{\makebox(0,0){$\markassin$}}}
\thicklines 
\path(461,90)(474,90)(487,91)(500,93)(513,95)(526,97)(539,100)(553,104)(566,108)(579,112)(592,117)(605,123)(618,129)(631,136)(645,143)(658,151)(671,159)(684,168)(697,177)(710,187)(723,197)(737,208)(750,219)(763,231)(776,243)
\thicklines 
\path(776,243)(789,256)(802,270)(815,283)(828,298)(842,313)(855,328)(868,344)(881,361)(894,378)(907,396)(920,414)(934,432)(947,451)(960,471)(973,491)(986,512)(999,533)(1012,555)(1026,577)(1039,600)(1052,623)(1065,647)(1078,672)(1091,697)(1104,722)
\end{picture}
    \caption{Computed spectrum of singularities: numerical ($+$) and
      theoretical (---) results.}
    \label{fig:mulfor}
  \end{center}
\end{figure}

We tested this algorithm on the process that we synthesized in
\S~\ref{sec:synth-mult-proc}, with the same parameters as on
Figure~\ref{fig:mulpro}, except that it was computed with $2^{22}$
points to get a sufficient scale range. The analyzing wavelet is
Daubechies 3 (extremal phase), different from the synthesizing
wavelet. Implementation is straightforward; we used a simple linear
regression on the 10 largest scales to compute $\lambda(\alpha)$, and
then $d_{2}(h)$.  Results are shown on Figure~\ref{fig:mulfor}.

\section{Generalized selfsimilarity}

The model for fully developed turbulence proposed in Castaing \emph{et
  al.} \cite{chilla96:_multip} asserts that the velocity field is a
random process $X$, with increments at scale $l$ following a law of
density $P_l$, and that if $l < L$,
\begin{equation}
  \label{eq:7}
  P_l(x) = \int g_{l L}(u) e^{-u} P_{L}(e^{-u} x) du,
\end{equation}
where the \emph{selfsimilarity kernel} satisfies $g_{l L} = g_{ll'}
* g_{l' L}$ for $l < l' < L$. This is a generalization of
the notion of selfsimilar process, because taking $g_{l L}\deq
\delta_{H \ln\frac{l}{L}}$ in~(\ref{eq:7}) yields
\begin{equation*}
  P_l(x) = \paren{\frac{L}{l}}^H P_L\paren{\paren{\frac{L}{l}}^H
    x} \Longleftrightarrow X_l \eqLaw \paren{\frac{l}{L}}^H X_L.
\end{equation*}

The construction of such a process for general $g$ is still an open
problem, but a discrete approach can be done using wavelets. Assuming
that for all $k$, $C_{j, k} \eqLaw X_{2^{-j}}$, and that, at $j$
fixed, the common probability density for $-\log_2(\abs{C_{j, k}})$ is
$\bsb{\tilde\rho}_j$, (\ref{eq:7}) becomes for $j > J$
\begin{equation}
  \label{eq:8}
  \bsb{\tilde\rho}_j = G_{j J} * \bsb{\tilde\rho}_J,
\end{equation}
where $G_{j J}(u) \deq g_{2^{-j} 2^{-J}}(-u)$ must satisfy
$G_{j J}=G_{j j'}*G_{j' J}$ for $j > j' > J$. One can furthermore assume
that $G_{j J}$ depends only on $j-J$, in which case $G_{j J} =
G^{*(j-J)}$.  Remark that, with the notation of
Definition~\ref{defi:rws}, $\bsb{\rho}_j(d \alpha) = j
\bsb{\tilde\rho}_j(j\alpha) d \alpha$. %
A RWS satisfying (\ref{eq:8}) can be obtained by simply taking
$\bsb\rho_j(d \alpha) = j G_{j0}(j\alpha) d \alpha$. Then its spectrum of
singularities can be computed almost surely and, thanks to
Theorem~\ref{theo:rwsspectrum}, it satisfies the multifractal
formalism given by (\ref{eq:5}), where the infimum is taken for ${q
  \geq q_c}$.
\paragraph{Example 1:}
$G = N_{m,\sigma^2}$, with $m > \sigma\sqrt{\frac{
    2}{\log_{2}(e)}}$. For all $\alpha$,
\begin{equation*}
  \bsb\rho(\alpha) = 1 - \log_{2}(e) \frac{(\alpha-m)^{2}}{2\sigma^2}.
\end{equation*}
The wavelet coefficients follow a log-normal law; the spectrum of
singularities is a segment of a parabola followed by a segment of
a line.

\paragraph{Example 2:}
$G = \delta_{\alpha_0} * \gamma_{\nu,\beta}$, with $\nu,\beta > 0$;
$\alpha_{0} > \alpha^{\star}(\nu,\beta)$, which is the largest
solution to $1 + \nu\log_2(-\alpha^{\star}) + \beta \log_2(e)
\alpha^{\star} + \nu \log_2\paren{\frac{\beta e}{\nu}} = 0$. For
$\alpha > \alpha_0$,
\begin{equation*}
  \bsb\rho(\alpha) = 1 + \nu\log_2(\alpha-\alpha_0) - \beta
  \log_2(e) (\alpha-\alpha_0) + \nu \log_2\paren{\frac{\beta
      e}{\nu}}.
\end{equation*}

\paragraph{Example 3:}
$G = \delta_{\alpha_{0}} * p_{c}$, where $p_{c}$ is a Poisson
distribution with parameter $c$ and $\alpha_{0} > \alpha^{\star}(c)$,
which is the solution to $1 - c \log_{2}(e) - \alpha^\star
\log_{2}\paren{\frac{c e}{ - \alpha^\star}} =0$. This kernel was
proposed by Dubrulle \cite{dubrulle} and She and Waymire
\cite{shewaymire} in the study of fully developed turbulence. For
$\alpha > \alpha_{0}$,
\begin{equation*}
  \bsb\rho(\alpha) = 1 - c \log_{2}(e) + (\alpha - \alpha_{0})
  \log_{2}\paren{\frac{c e}{\alpha - \alpha_{0}}}
\end{equation*}

It is remarkable that such selfsimilarity kernels can be obtained in
the framework of RWS, since they are usually expected to be the
signature of cascade models on the wavelet coefficients which display
strong correlations between these coefficients (see \cite{BacMuz}).
Our examples show that it is not the case: Correlations between
wavelet coefficients cannot be inferred from the particular shape of
the p.d.f. of the wavelet coefficients at each scale.  Note however
that one possible option, in order to derive some information on these
correlations, is to study how statistics of local suprema of the
wavelet coefficients behave, see \cite{Jaff3}.

\bibliographystyle{abbrv}
\bibliography{Auteurs,mrabbrev,Editeurs,Math,Phys,Preprints}

\end{document}